\documentclass[tighten,twocolumn]{aastex62}
\usepackage[]{graphicx}
\usepackage{amsmath}
\usepackage[normalem]{ulem}
\usepackage{lineno}

\newcommand{\changed}[1]{#1}

\shortauthors{KERR}
\shorttitle{Single Pulse Variability in Gamma-ray Pulsars}

\begin{document}

\title{Single Pulse Variability in Gamma-ray Pulsars}

\author[0000-0002-0893-4073]{M.~Kerr}
\affiliation{Space Science Division, Naval Research Laboratory, Washington, DC 20375--5352, USA}
\correspondingauthor{M.~Kerr}
\email{matthew.kerr@nrl.navy.mil}

\begin{abstract} The \textit{Fermi} Large Area Telescope receives
  $\ll$1 photon per rotation from any $\gamma$-ray pulsar. However,
  out of the billions of monitored rotations of the bright
  pulsars Vela (PSR~J0835$-$4510) and Geminga (PSR~J0633$+$1746), a
  few thousand have $\geq$2 pulsed photons.  These rare pairs encode
  information about the variability of pulse amplitude and shape.  We
  have cataloged such pairs and find the observed number to be in good
  agreement with simple Poisson statistics, limiting any amplitude
  variations to $<$19\% (Vela) and $<$22\% (Geminga) at 2$\sigma$
  confidence.  Using an array of basis functions to model pulse shape
  variability, the observed pulse phase distribution of the pairs
  limits the scale of pulse shape variations of Vela to $<$13\% while
  for Geminga we find a hint of $\sim$20\% single-pulse shape
  variability most associated with the pulse peaks.  If variations
  last longer than a single rotation, more pairs can be collected,
  and we have calculated upper limits on amplitude and shape
  variations for assumed coherence times up to 100 rotations, finding
  limits of $\sim$1\% (amplitude) and $\sim$3\% (shape) for both
  pulsars.  Because a large volume of the pulsar magnetosphere
  contributes to $\gamma$-ray pulse production, we conclude that the
  magnetospheres of these two energetic pulsars are stable over one
  rotation and very stable on longer time scales.  All other
  $\gamma$-ray pulsars are too faint for similar analyses.  These
  results provide useful constraints on rapidly-improving simulations
  of pulsar magnetospheres, which have revealed a variety of
large-scale instabilities in the thin equatorial current sheets where
the bulk of GeV $\gamma$-ray emission is thought to originate.
\vspace{1cm} \end{abstract}


\section{Introduction}
\label{sec:intro}

Our understanding of pulsar magnetospheres is improving rapidly:
sophisticated simulations  are providing the first nearly
self-consistent realizations of magnetospheres, and the \textit{Fermi}
Large Area Telescope (LAT) continues its groundbreaking discovery and
characterization of the large population of $\gamma$-ray pulsars
needed to test these models.

These new particle-in-cell (PIC) simulations rest on half a century of
theoretical work.  Soon after \citet{Hewish68} discovered pulsars,
\citet{Goldreich69} proposed a ``force-free'' (FF) model of the pulsar
magnetosphere in which a co-rotating charge-separated plasma with
density $\rho_{\mathrm{GJ}}\sim\Omega\cdot B$ screens the parallel electric
field ($E_\parallel$), a picture which largely holds up today.
However, FF magnetospheres by definition cannot accelerate particles,
and the necessity of strong acceleration with concomitant $e^+$/$e^-$
pair formation in active pulsars was quickly recognized
\citep{Sturrock71}.  Moreover, solving the coupled currents and
electrodynamics even with the FF constraint proved intractable, and
the best available analytic model was that of a magnetic dipole
rotating \textit{in vacuo} \citep{Deutsch55}, also clearly far from
reality.

These difficulties led to long threads of research which
compartmentalized the problem.  To characterize particle acceleration,
theorists supposed that in an otherwise FF magnetosphere, small
regions might lack the $\rho=\rho_{\mathrm{GJ}}$ plasma needed to
screen the electric field.  Such vacuum ``gaps'' could accelerate
particles to high energy and form the pairs needed to achieve the
assumed force-free conditions elsewhere.  The ideas included a gap
operating above the magnetic polar cap \citep{Ruderman75}, a ``slot
gap'' arising around the polar cap rim \citep{Arons83}, and an ``outer
gap'' operating above the null charge surface ($\rho_{\mathrm{GJ}}=0$)
near the light cylinder \citep{Cheng86}.

$\gamma$ rays are ideal for testing these models, as they directly
trace particle acceleration.  Observations by COS-B and EGRET
\citep[e.g.][]{Thompson99} provided both pulse profiles and spectra
for a half-dozen young pulsars.   Even in small numbers, it was clear
that the light curves differed substantially from radio pulses,
arriving later in the neutron star rotation (by a phase lag commonly
notated $\delta$) and demonstrating a now-canonical morphology of two
cuspy peaks separated by $\Delta$$\sim$0.2--0.5 rotations.  Comparison
to data was achieved using a ``geometric'' method which propagated
$\gamma$-ray emissitivities to an observer using the Deutsch field.
Approximately offsetting effects from time delay and field sweepback
produced caustic-like features \citep{Romani95} more reminiscent of
the data than the predictions of low-altitude models
\citep{Daugherty96}, lending some credence to an outer magnetospheric
origin.  However, the small pulsar numbers and incomplete magnetic
field information could not support strong conclusions.

Thirty years after the proposal of the idea, \citet{Contopoulos99} and
\citet{Timokhin06} numerically solved the FF magnetohydrodynamic
equations for an aligned ($\Omega\parallel B$) magnetosphere while
later work \citep{Spitkovsky06,Kalapotharakos09} addressed the
nonaligned cases.  These breakthrough solutions finally provided an
exact description of both fields and currents, revealing a key
feature: current flows out of the polar cap into a broad volume of the
magnetosphere, but \changed{much of it} returns in a very thin layer
in the equatorial plane, splitting at the ``separatrix'' near the
light cylinder and returning to the rim of the polar cap along the
boundary of the closed and open magnetic field zones.  This equatorial
current sheet (ECS) is a hallmark of FF magnetospheres,
\changed{though less of the return current flows through the ECS for
large magnetic inclinations.}  Its extremely
high areal current density makes it a natural candidate for magnetic
reconnection and accelerating fields, but assessing these processes
cannot be done with FF MHD.

This development set the stage for an observational revolution: soon
after its launch, the \textit{Fermi} Large Area Telescope
\citep[LAT,][]{Atwood09} detected populations of radio-loud,
radio-quiet \citep{Abdo09_sciblind}, and millisecond pulsars
\citep{Abdo09_scimsp}, establishing all classes as efficient
$\gamma$-ray emitters capable of converting $\sim$1--30\% of their
available spin-down power \citep{Abdo13_2PC} into GeV photons. Early
analyses of the LAT data strongly supported the outer magnetosphere as
the most likely site of substantial $\gamma$-ray production.  First,
the spectral signatures of bright pulsars lacked evidence for
attenuation in the strong magnetic fields above polar caps
\citep{Abdo09_vela}.  Second, the fraction of pulsars with
radio-trailing, cuspy double-peaked light curves was too large for
low-altitude emission models \citep[e.g.][]{Watters09}.  However, the
expanding pulsar population soon challenged the simple geometric
prescriptions, and e.g.  \citet{Pierbattista15} found in general that
radio-loud $\gamma$-ray pulsars cannot be consistently modeled with
any combination of a ``gap'' accelerator and the vacuum magnetic field.

\citet{Bai10} reinforced this discrepancy, finding that replacing the
Deutsch field with a FF magnetosphere yielded poor agreement with
\textit{Fermi} data when paired with existing gap models, largely due
to the increased field sweepback.  Instead, these authors invoked a
``separatrix layer model'' operating along the ECS,
extending even further into the outer magnetosphere than the classic
outer gap model.  Relaxing the FF assumptions by introducing
resistivity into the magnetosphere---thus emulating conversion of the
fields and currents into high-energy particles---yielded even better
agreement.  \citet{Kalapotharakos14} used a prescription which was FF
inside the light cylinder and resistive outside, with the resulting
tweaks to the magnetic field and emission geometry producing the first
excellent agreement with the observed $\delta$-$\Delta$ properties of
the LAT pulsar sample \citep{Abdo13_2PC}.  This and other work
cemented the paradigm that $\gamma$-ray pulsars have nearly force-free
magnetospheres and produce most of their $\gamma$ rays in the ECS near
and beyond the light cylinder.

Recent PIC simulations have confirmed this picture with, for the first
time, nearly self-consistent modeling.  In PIC simulations,
macroparticles representing collections of many charged particles are
evolved iteratively with the electromagnetic fields tabulated at the
cell boundaries.  The particles produce the (non background) fields
and are in turn accelerated by them.  The dynamic range of the
simulations is still many orders of magnitude below a true
magnetosphere, so e.g. maximum particle Lorentz factors reach
$\sim$1000.  Thus the key process of pair production must still be put
in manually.  \citet{Timokhin13} demonstrated that a current density
$J\equiv c \rho$ of $J/J_{\mathrm{GJ}}<0$ or $J/J_{\mathrm{GJ}}>1$ is
indicative of intense particle acceleration and pair production, while
for $0<J/J_{\mathrm{GJ}}<1$, current can be sustained by space-charge
limited flow with no pair production.  This provides a common
prescription for PIC simulations: by requiring that
$J/J_{\mathrm{GJ}}>1$, authors can assume pair production occurs
locally\footnote{Simulations are often scaled to $\rho_{\mathrm{GJ}}$,
so this condition also appears in the literature as a magnetization
exceeding unity.} and replace the primary particles with secondaries
and $\gamma$ rays, thus providing both the copious plasma needed to
screen the electric field and tracking the emission of $\gamma$ rays
for comparison to observation.

With such prescribed pair injection, PIC simulations have been
successful in reproducing the general structure of FF magnetospheres
but offer more insight into the dynamics, in particular the response
of the magnetosphere to the location and extent of pair production.
E.g., \citet{Philippov15b} found that the incorporation of General
Relativistic effects near the neutron star surface suppresses
$\rho_{\mathrm{GJ}}$ locally, facilitating the pair production needed to
explain the presence of radio pulsations.  However, \citet[][and see
also \citet{Kalapotharakos18} for similar studies]{Chen20}, simulating
magnetospheres over a range of pair production rates, highlight the
importance of pairs within the ECS to achieving a near-FF
magnetosphere due to the difficulty in screening $E_{\parallel}$
there.  If pairs produced elsewhere \citep{Cerutti15, Brambilla18}
\changed {are unable to} enter the ECS, or the sheet itself is no
longer able to support pair production, the magnetosphere may
transition to a static, charge-separated electrosphere. In the
intermediate regime, the magnetosphere fluctuates between these
states, with $\gamma$-ray production and global currents fluctuating
in step.

As PIC simulations have been refined, they have become more predictive
for $\gamma$-ray data.  \citet{Kalapotharakos18} and
\citet{Philippov18} built $\gamma$-ray light curves by tracking
emitted photons and found their pattern on the sky in reasonable
agreement with data.  \changed{\citet{Philippov18}} postulate that $\gamma$ rays
are primarily produced by synchrotron emission from reconnection in
the ECS \citep{Uzdensky14}, near the LC in the case of aligned
rotators and from beyond the LC for oblique rotators, which are also
somewhat less efficient emitters.  On the other hand,
\citet{Kalapotharakos17} and \citet{Kalapotharakos19} have noted that
$\gamma$-ray properties lie on a ``fundamental plane'' according to
their $\gamma$-ray luminosity, spectral cutoff energy, magnetic field,
and spindown luminosity $\dot{E}$.  By connecting the cutoff energy to
the maximum particle Lorentz factors, they argue that the scaling is
more indicative of curvature radiation.  The upcoming third
\textit{Fermi} pulsar catalog will provide data for many more pulsars
and provide a touchstone to determine which of these mechanisms is
operant.  Thus, after more than 50 years of work, theory and data are
poised to deliver a nearly-complete picture of the pulsar
magnetosphere and how it emits $\gamma$ rays.

Yet while the overall static picture is becoming clear, the PIC
simulations have opened the field of magnetosphere dynamics and their
observability.  Notably, kink instabilities in the ECS are a common feature in PIC simulation results---especially near
the Y-point/separatrix where the return current converges
\citep{Chen14}---with the local
tangent to the ECS varying by $\sim$0.1 rad or more, which could
affect the beaming of emitted $\gamma$ rays.  If the $\gamma$ rays are
emitted via magnetic reconnection, \citet{Uzdensky14} argue that
the maximum size of coherent regions should be $<$100$\times$ the width of
the current sheet, only about 1\,m!  Thus many reconnecting
``islands'' contribute to the current sheet and this could suppress such
coherent effects, whereas dynamic instabilities could affect curvature
radiation emission more significantly.  On the other hand,
reconnecting regions could merge into larger ones, driving
$\gamma$-ray variability \citep{Philippov19,CeyhunAndac22}. Further, \citet{Kirk17} invoke a bulk Lorentz factor of $\sim$10$^4$ in the wind of the Crab pulsar, in which case reconnection cannot proceed due to time dilation.  In any case, for weaker pulsars, transitions
between FF magnetospheres and ``dead'' electrospheres are common
\citep{Kalapotharakos18,Philippov18} and
transitions can happen on timescales of one neutron star rotation
\citep{Chen20} and longer.

$\gamma$-ray pulses, tracing both of these processes, would provide an
invaluable probe for dynamics of the equatorial current sheet and the
global magnetosphere.  Accordingly, this work presents the first
systematic search for and characterization of single- and few-pulse
$\gamma$-ray pulse variability.  Despite the recent theoretical
motivation, such studies have heretofore been lacking primarily due to
observational constraints: even the sensitive LAT records on average
only $\sim$0.003 photons from Vela, the brightest $\gamma$-ray, during
each 89\,ms rotation, so measuring pulse-to-pulse profile or amplitude
is impossible.

However, the LAT has observed billions of rotations of Vela, and
$\sim$$10^{-6}$ of the rotations have 2 or more recorded photons.  These
multi-photon events are a far cry from a pulse, yet
they still carry information about the variability of the pulsed GeV
emission on short (neutron star rotation) time scales.  For instance,
any deviations from a constant flux would inevitably increase the
number of observed pairs over the simple Poisson prediction.
Likewise, pulse shape variations can be probed by examining the joint
distribution of pulse phase for photon pairs.  If certain portions of
the pulse profile are brighter during some rotations than in others,
this will manifest as a correlation in phase.
Most generally, we expect simultaneous amplitude and shape variations:
e.g., in a two-peaked pulse profile, a reduction in the intensity of
one of the peaks could entail a reduction in total intensity, or it
could be accompanied by an offsetting increase in the brightness of
the other peak.

Below, we carry out a search for single- and mulitple-pulse
variations in the two brightest $\gamma$-ray pulsars, Vela and
Geminga.  \S\ref{sec:data} presents our data preparation and selection
of ``photon pairs''.  In \S\ref{sec:amplitude}, we search for
amplitude variations independently and (finding none), we use the
resulting limits as guidelines for a pulse
shape variation search (\S\ref{sec:pulse_shape}) in which the total flux doesn't vary by more
than the established limits.
We extend the analysis to longer timescales (up to 100 rotations) in
\S\ref{sec:multirot}, improving the constraints.  We discuss other
$\gamma$-ray pulsars, which are too faint to be of use, in
\S\ref{sec:other}, and we conclude in \S\ref{sec:discussion}, placing
our results in context and discussing the scant prospects for
additional $\gamma$-ray variability measurements.  \footnote{The
software developed for this analysis is archived at: https://doi.org/10.5281/zenodo.6636586.}

\begin{figure}
\centering
\includegraphics[angle=0,width=0.98\linewidth]{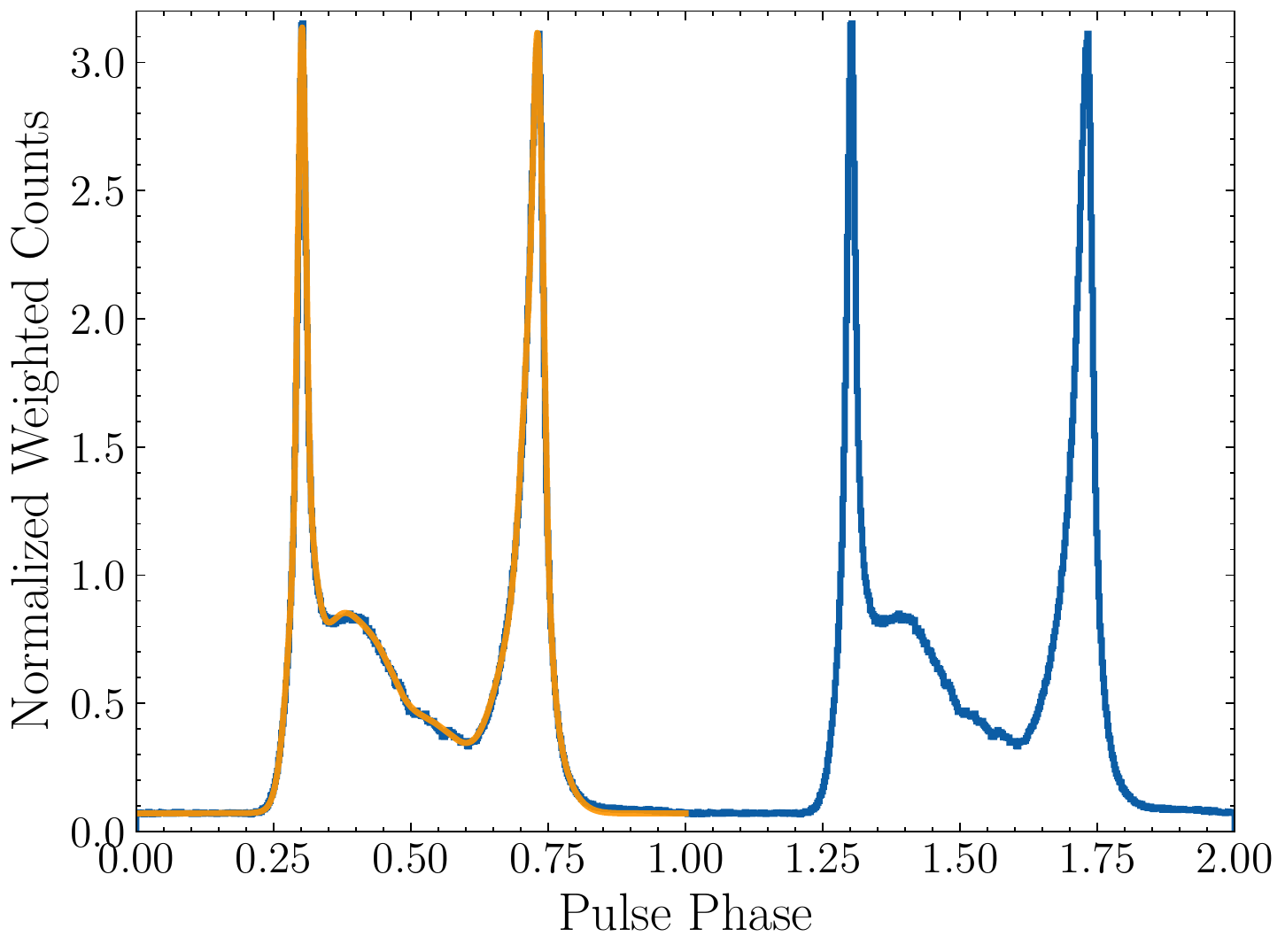}
\caption{\label{fig:vela_prof}Two rotations of the Vela pulsar,
showing a weighted histogram (500 bins per rotation, blue) of the full dataset and an
  analytic model for $f(\phi)$ overlaid (orange).}
\end{figure}

\section{Data Preparation and Selection}
\label{sec:data}

We obtain ``Pass 8'' \citep{Atwood13} event reconstruction version
P8R3 \citep{Bruel18} from the \textit{Fermi} Science Support
Center\footnote{https://fermi.gsfc.nasa.gov/ssc/data/access/} and
process it using the \textit{Fermi} Science Tools v2.0.0.  We select
about 12 years of data, between 2008 Aug 04 (MJD 54698) and an end
date which depends on the range of validity of the pulsar timing
solution (Table \ref{tab:data}).  We further restrict attention to
events with reconstructed energy 0.1\,GeV$<$E$<$10\,GeV, measured
zenith angle $<$100$^\circ$, and a reconstructed incidence direction
placing the photon within 3$^\circ$ of the pulsar position.  To
enhance the sensitivity of the analyses \citep{Bickel08, Kerr11,
Kerr19}, using the 4FGL DR2 sky model \citep{4FGL, 4FGL_DR2} and the
\texttt{P8R3\_SOURCE\_V3} instrument response function, we
assign each photon a weight using \texttt{gtsrcprob} and restrict
attention to events with $w>0.05$.  This selection retains most of the
signal while eliminating background photons that increase
computational costs.

\begin{table}
\centering
\begin{tabular}{l | l | l | l }
Name & Nickname & MJD End  & Years \\
\hline
\hline
J0835$-$4510 & Vela & 59170 & 12.3 \\
J0633$+$1746 & Geminga & 59249 & 12.5 \\
J0534$+$2200 & Crab & 59000 & 11.9 \\
J1709$-$4429 & -- & 59312 & 12.6 \\
J1057$-$5226 & -- & 59249 & 12.5  \\
\hline
\end{tabular} 
\caption{\label{tab:data}Some basic properties of the target sources
  and the data set.  All data sets begin MJD 54698.}
\end{table}

\begin{table}
\centering
\begin{tabular}{l | r | r | r | r}
Type & Obs. & Chance  & \textit{bona fide}  & Pred.\\
\hline
\hline
  Vela \hspace{0.5cm}  \\
  \hline
  \dotfill events & 3,295,429 & --- & 2,720,089.2 & --- \\
  \dotfill pairs & 6,388 & 2,024.4 & 4,363.6 & 4,295.4 \\
  \dotfill triples & 12 & 6.6 & 5.4 & 4.7 \\
  \hline
  Geminga   \\
  \hline
  \dotfill events & 1,154,145 & --- & 980,774.8 & --- \\
  \dotfill pairs & 1,906 & 527.9 & 1,378.1 & 1,417.0 \\
  \dotfill triples & 5 & 2.4 & 2.6 & 1.5 \\
\hline
\end{tabular} 
\caption{\label{tab:counts}Observed total events and probability-weighted pairs and triples.  The
first column gives the total number of coincidences.  The second
tabulates the expected number of false positives (1 or more background
events).  The third column gives the prediction for the total number
of true pairs (triples) based on the photon weights, while the fourth
gives the prediction from the no-variability hypothesis, i.e. folding
a constant rate through the instrument response function (see main
text).}
\end{table}

We obtain timing solutions for each pulsar using the maximum
likelihood approach presented in \citet{GWB22}.  Using PINT
\citep{luo21}, we computed the absolute pulse phase for each event,
and we define \textit{pairs} as those photons with identical integer
phase.  This definition is somewhat arbitrary, but we have centered
the pulse profile (Figure \ref{fig:vela_prof}) within the phase window
such that pairs associated with \textit{the same pulse} are selected.
(In \S\ref{sec:multirot}, we adopt a simpler method of selecting
photons whose total phase difference is simply less than some
threshold, and we find no substantive difference.)  \textit{Triples}
and higher-order coincidences follow the same definition.

Pursuing this method for Vela, we identified 6388 pairs, 12 triples,
and no higher order coincidences.  Because some photons come from the
background, this total comprises three cases: two background photons,
a single pulsed photon with a background photon, and two \textit{bona
fide} pulsed photons.  The breakdown can be estimated from the photon
weights, where $w_1$ and $w_2$ are the probability weights of the two
photons in the observed pair: $\sum_i (1-w_{1i})(1-w_{2i})=188.9$
background pairs; $\sum_i (1-w_{1i})w_{2i} + w_{1i}(1-w_{2i})=1835.4$
with a single pulsed photon; and $\sum_i w_{1i}w_{2i}=4363.6$
\textit{bona fide} pairs.

Similar arguments apply to the triples and we expect $\sum_i
w_{1i}w_{2i}w_{3i}=5.4$ \textit{bona fide} triples.  These triples are
too rare to be significantly constraining, so in the analysis below we
focus on pairs.  Table \ref{tab:counts} summarizes these results for
Vela and Geminga,
including predictions accounting for the instrument exposure (see below).

\section{Amplitude Variations}
\label{sec:amplitude}

\begin{figure}
\centering
\includegraphics[angle=0,width=0.98\linewidth]{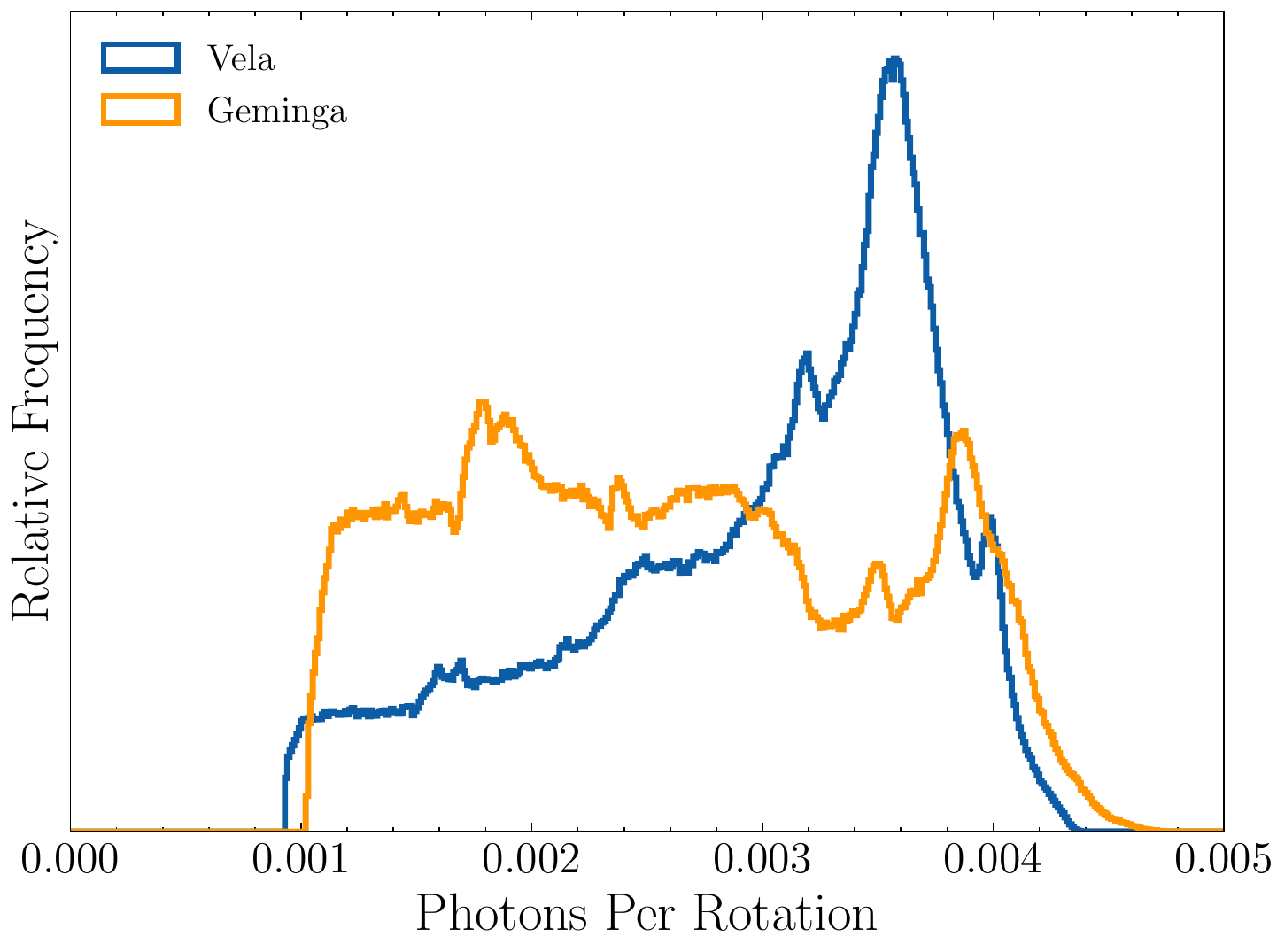}
  \caption{\label{fig:vela_geminga_rate_comp}The distribution of the
  expected number of photons per rotation, averaged over a 30-s
  interval, for Vela and Geminga.  The largest values
  correspond to the pulsar nearing the center of the field-of-view,
  and Vela accumulates more exposure under these favorable circumstances, enhancing the number of photon pairs expected relative to Geminga.}
\end{figure}

We test for amplitude variations against the null hypothesis---no
variations---in which case the photon arrivals follow a Poisson
distribution.  To do this, we compute the expected number of photons
for each rotation of the pulsar over the entire 12-year dataset
by folding an average spectral energy distribution through the
\textit{Fermi} instrument response function.

In practice, we can ignore any intervals during which the pulsar is
unobserved, as these have no constraining power.  \changed{Moreover,
the apparent angular motion of any source in the \textit{Fermi}-LAT
field-of-view over 30\,s is $<$2$^{\circ}$, allowing the exposure to
be computed at this cadence with $<$0.1\% error.  So as} to
simplify the calculation, we determine the instantaneous photon rate
over these 30\,s intervals and assume it is the same for each pulsar
rotation.  As a further simplification, we ignore the spin evolution
of the pulsar and simply use the mean spin rates of
$\bar{\nu}=11.187519$\,Hz for Vela and $\bar{\nu}=4.217539$\,Hz for
Geminga.  With this prescription, the expected number of source
photons per rotation is given by
\begin{equation}
  \lambda_{30} = \frac{1}{\bar{\nu}}\int N(E) \epsilon(E,\cos(\theta_{30}))\,dE,
\end{equation}
with $N(E)$ the photon spectral density
(ph\,cm$^{-2}$\,s$^{-1}$) for the pulsar and $\epsilon$ the
effective area (cm$^2$) for the given energy $E$ at an incidence angle
$\theta_{30}$ taken to be constant over the 30\,s interval.  We evaluate
$\lambda_{30}$ using the methods developed by \citet{Kerr19},
particularly the package
\texttt{godot}\footnote{https://www.github.com/kerrm/godot}.  In brief,
this evaluation uses the tabulated spacecraft pointing history (``FT2
file''), with 30-s resolution, to determine $\cos\theta_{30}$ and
hence the appropriate value of the tabulated effective area.  Gaps in
data taking are accounted for with the tabulated ``Good Time
Intervals'' (GTI).  The method also assumes \changed{a universal
energy weighting $\propto$$E^{-2}$, which roughly describes the
overall distribution of LAT photon energies.  To further tune this
approach for pulsars, which have spectral cutoffs at a few GeV, we
perform the exposure calculation using 8 logarithmic bins spanning
100\,MeV to 10\,GeV.  In each band, we then estimate the constant of
proportionality to enforce $\sum_i \lambda_{30}^i=\sum_i
w_i/30\mathrm{s}$.  This accounts for any ``local'' differences in the
distribution of photons from $\propto$$E^{-2}$ as well as for the
reduction in effective exposure at low energies because the point
spread function (PSF) is broader than the 3$^\circ$ region of
interest.  We further note that we tabulate events which convert in
the ``front'' (\texttt{evtype} 1) and ``back'' (\texttt{evtype} 2) of
the tracker separately.}
We
exclude any intervals with $\cos(\theta_{30})<0.4$, when the source is
near or outside the edge of the LAT field-of-view.  \changed{We
further discard the lowest 0.5\% of intervals, which eliminates a tail
of low-exposure instances.}  The resulting rates (per rotation) for the
two pulsars are shown in Figure \ref{fig:vela_geminga_rate_comp}.
It is interesting to note that despite similar median photon rates
(0.0032 and \changed{0.0025} per rotation), Vela has three times as many photon
pairs.  As Figure \ref{fig:vela_geminga_rate_comp} shows, Vela
accumulates much more ``high rate'' exposure
than Geminga, and this leads to more pair production.
This intuitive explanation is borne out by the
computations below and illustrates the importance of correcting for
the instantaneous instrument exposure.

With the computed rates, we can predict the total number of pairs
expected to be observed by using the Poisson distribution to evaluate
the probability mass function of observing 0, 1, 2, or 3 photons
during each rotation and summing these probabilities over all observed
pulsar rotations.  Using a simple model in which the pulsed flux is
constant and only the instrumental exposure varies, for Vela this
process yields \changed{$4,295.4 \pm 65.5$ pairs (uncertainty assumes Poisson
variance), in good agreement (1.0$\sigma$) with the observed, probability-weighted
value of 4,363.6.  Although triples are too rare to be constraining,
the predicted number is 4.7, again in good agreement with the
observation of 5.4.  For Geminga, the prediction ($1,417.0\pm37.6$) is higher
than the observed value (1,378.1) and agrees at 1.0$\sigma$.  All
results are tabulated in Table
\ref{tab:counts}.}

\begin{figure}
\centering
\includegraphics[angle=0,width=0.98\linewidth]{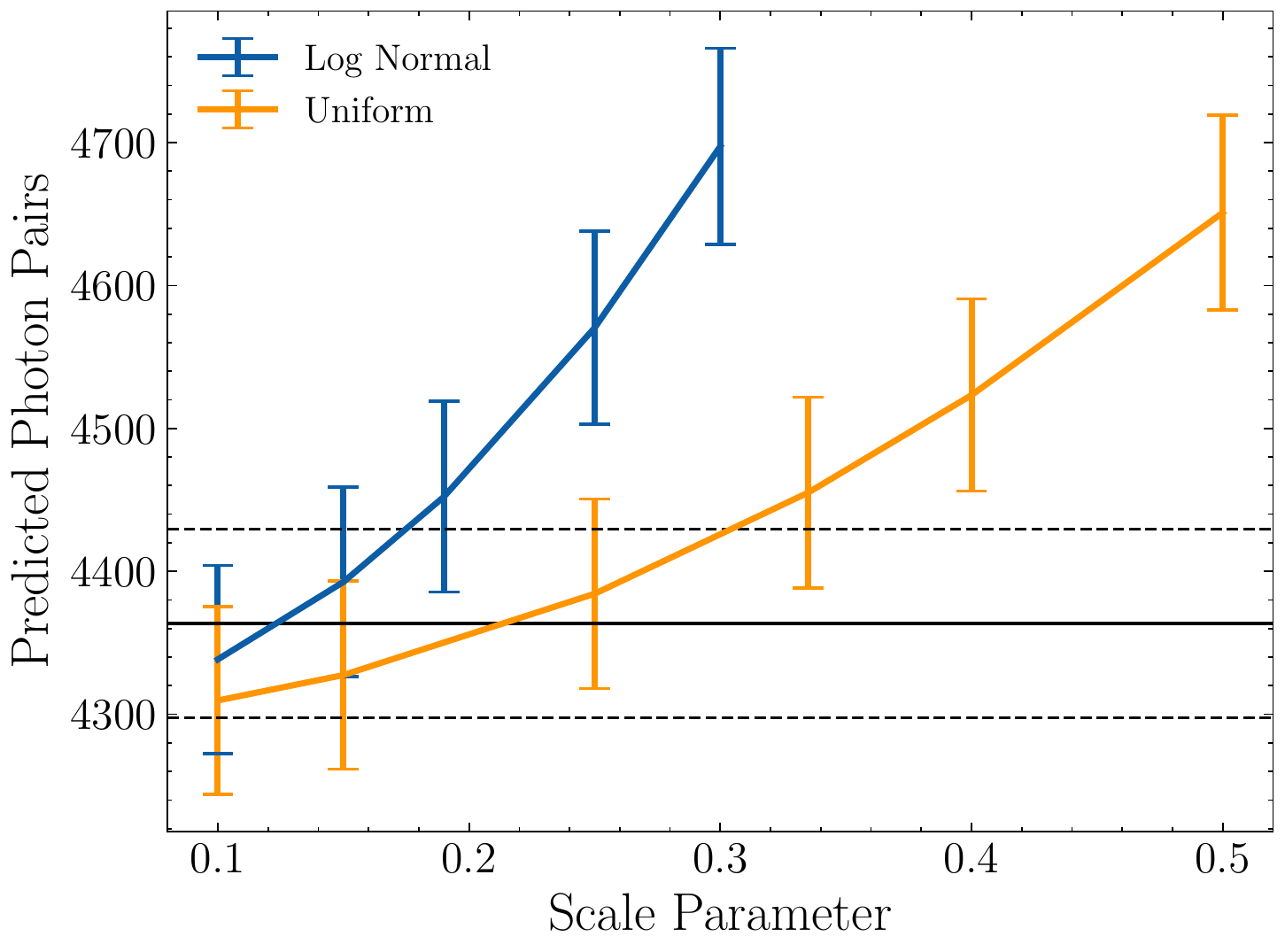}
\caption{\label{fig:vela_amp_model_comp}A comparison of the predicted count
of \textit{bona fide} photon pairs for two models along with the observed value (black line).  The scale parameter corresponds to $\sigma$ in the case
  of the log normal model and width $W$ in the uniform model.  The dashed lines indicate the 1$\sigma$ (Poisson) uncertainty range.
See main text for details.}
\end{figure}

\begin{figure*}
\centering
\includegraphics[angle=0,width=0.98\linewidth]{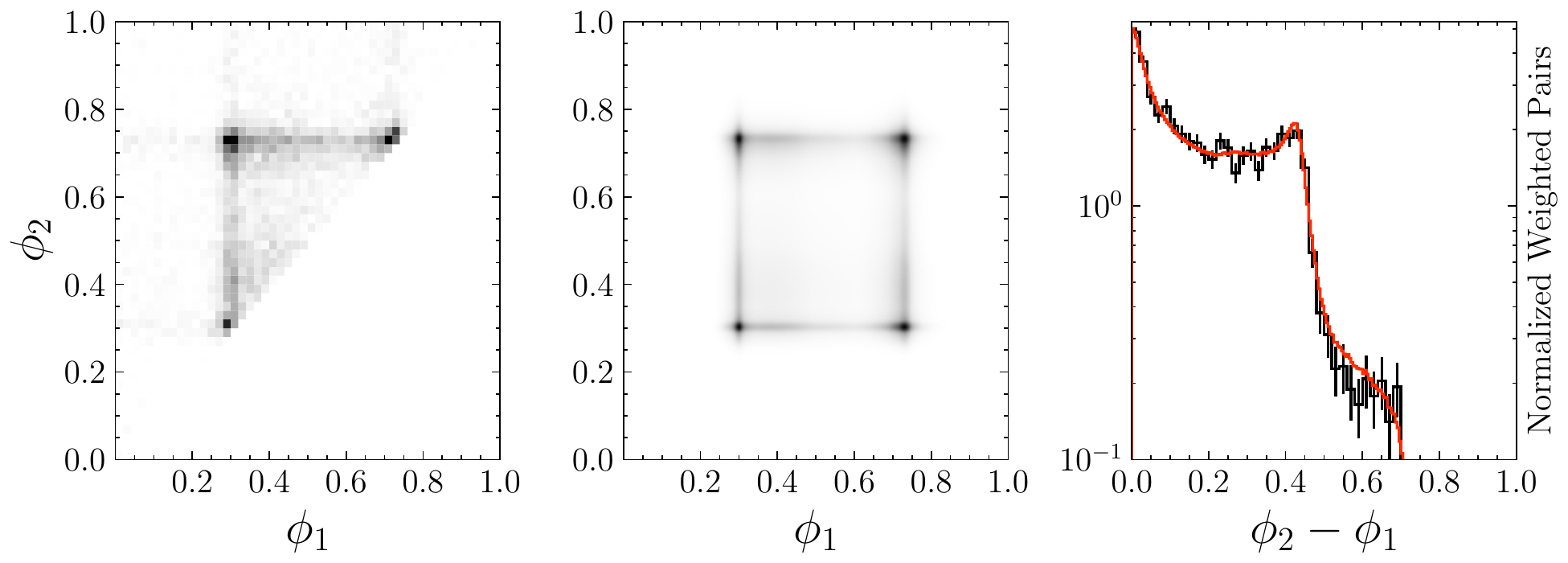}
\caption{\label{fig:vela_empirical_dphi}The joint distribution of
phase for photon pairs, weighted by $w_1\times w_2$.  The left panel
shows the observed distribution sorted such that $\phi_1<\phi_2$.  The
center panel shows the expected null distrubtion
$f(\phi_1,\phi_2)=f(\phi_1)f(\phi_2)$.  The right panel gives the
distribution of the sorted phase differences for the observed pairs,
again weighted by $w_1 \times w_2$, along with the expection from the
null distribution, obtained by randomizing phases from the entire
sample.  Error bars are computed from the Poisson variance, which neglects the small correlations between bins.}
\end{figure*}

\begin{figure*}
\centering
\includegraphics[angle=0,width=0.98\linewidth]{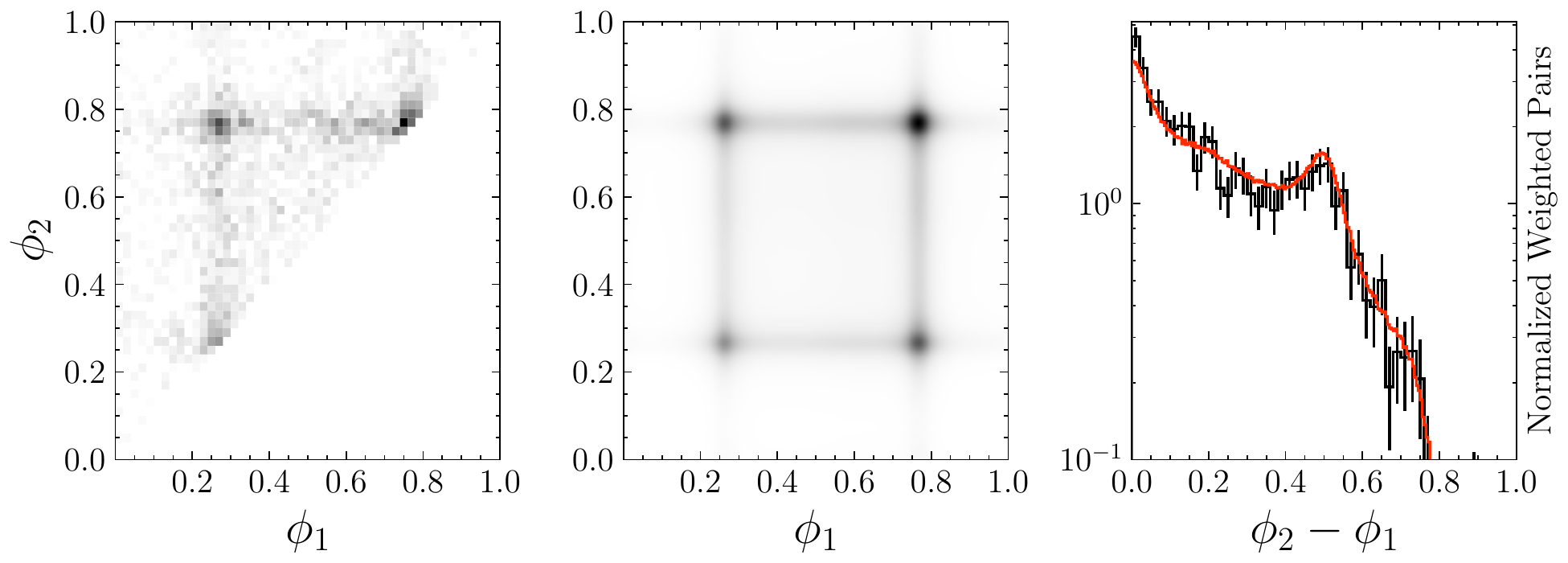}
\caption{\label{fig:geminga_empirical_dphi}As Figure
\ref{fig:vela_empirical_dphi} for Geminga.}
\end{figure*}

To gauge how large intrinsic flux variations could be without entering
serious disagreement with the pair count, we consider two models of
amplitude variations.  First, we use a log normal distribution, with
parameters $\mu$ and $\sigma$ for the underlying normal distribution.
If we fix $\mu=-\sigma^2/2$, the distribution has unit mean, which
preserves a constant photon rate but increases fluctuations between
pulses.  We then scale each value of $\lambda_{30}$ with a random draw
from the distribution.  (It is not necessary to draw a random variable for
every pulsar rotation because the distribution is already sufficiently
sampled by the millions of values of $\lambda_{30}$.)  Second, we
consider a uniform distribution with total width $2W$, which allows
the relative values of $\lambda_{30}$ to vary by up to $1+W$, and we
restrict $W\leq1$ to avoid negative photon rates.  These distributions
differ strongly: in particular, the log normal distribution has heavy
tails, allowing for very bright pulses, while the brightest pulses are
capped in the uniform case.

The results of the study for Vela are shown in Figure
\ref{fig:vela_amp_model_comp}.  Increasing the scale parameters
$\sigma$ and $W$ increases the number of predicted pairs until the
predictions are no longer compatible with the observed value.  If we
adopt a 2$\sigma$ discrepancy for an upper limit, then scale
parameters larger \changed{than $\sigma=0.21$ and $W=0.38$ are
excluded.}
\changed{To put the models on equal footing we can compare their
standard deviation ($\sigma_{w}=W/\sqrt{3}$ for the uniform case) and
find $\sigma\approx\sigma_w$, and so in a relatively model
independent fashion, we can exclude fluctuations whose typical strength
exceeds 21\%.  A similar analysis for Geminga gives limits of
$\sigma=0.16$ and $W=0.28$, limiting fluctuations to about 16\%.}

\changed{ These results depend on the observed values, and so Geminga
gives superior results despite having fewer pairs because the null
pair prediction exceeds the observed value.  We can instead simply
consider the capability of the data to constrain variations without
regard for the particular realization of it.  With $\sim$4300
predicted (null hypothesis) pairs, the largest fractional increase in
pair rate is $2\times4300^{-\frac{1}{2}}$=3.0\%, and for Geminga it is
$2\times1400^{-\frac{1}{2}}$=5.3\%.  These values correspond to
$\sigma$ and $W$ parameters of 0.17 and 0.30 for Vela, and for Geminga
0.23 and 0.40, or relative variations of 17\% and and 23\%
respectively.}

\section{Pulse Shape Variations}
\label{sec:pulse_shape}

We now consider limits on the variation in pulse shape.  Let the
time-averaged pulse profile be $f(\phi)$.  If normalized such that
$\int_0^1 f(\phi)=1$, this also is the probability density function (pdf)
to observe a photon at the given phase.  The pdf for the photon pairs
is $f(\phi_1,\phi_2)$, normalized in the same way, and in the null
hypothesis $\phi_1$ and $\phi_2$ are independent variables
$f(\phi_1,\phi_2) = f(\phi_1)f(\phi_2)$.  We can analyze any pairwise
pulse shape variations using this distribution.

First, we characterize the distribution empirically.  Figure
\ref{fig:vela_empirical_dphi} shows the observed joint probability
distribution $f(\phi_1,\phi_2)$ for Vela as estimated with a weighted
histogram, and the null expectation, namely $f(\phi_1)f(\phi_2)$.  The
data and this null model are visually in good agreement,
though the small counts preclude assessment of goodness of
fit.  We have also examined the distribution of $\phi_2-\phi_1$,
which also encodes information about correlations and is more
sensitive to fluctuations with a constant scale $\delta\phi$.  E.g.,
an enhancement of probability to observe a photon pair from
\emph{either} peak should produce an excess near the origin and a
deficit elsewhere.  This distribution, likewise, agrees well
(within the uncertainty estimates) with the null hypothesis
expectation, which we estimated by drawing random pairs from the
much larger parent sample.

\begin{figure}
\centering
\includegraphics[angle=0,width=0.98\linewidth]{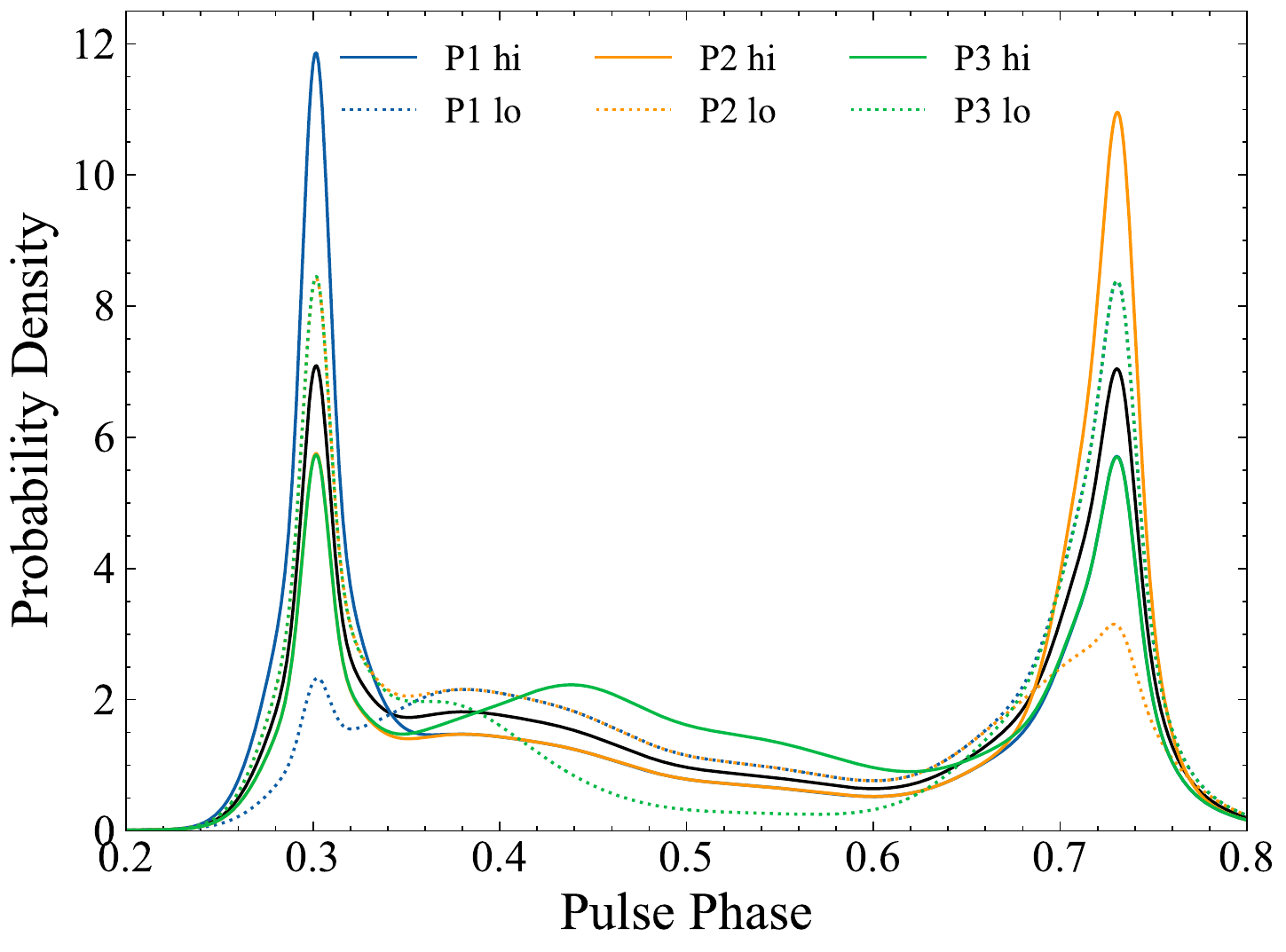}
\caption{\label{fig:vela_basis}An example set of six basis functions for
a model of single-pulse variations for Vela.  They correspond to an
addition (or loss) of 15\% of the total flux from each of the three
major components, P1, P2, and P3/the bridge.  The time-averaged
pulse profile is shown in black for reference.  By construction, the
mean of the six basis functions is equal to this pulse profile.}
\end{figure}

\begin{figure}
\centering
\includegraphics[angle=0,width=0.98\linewidth]{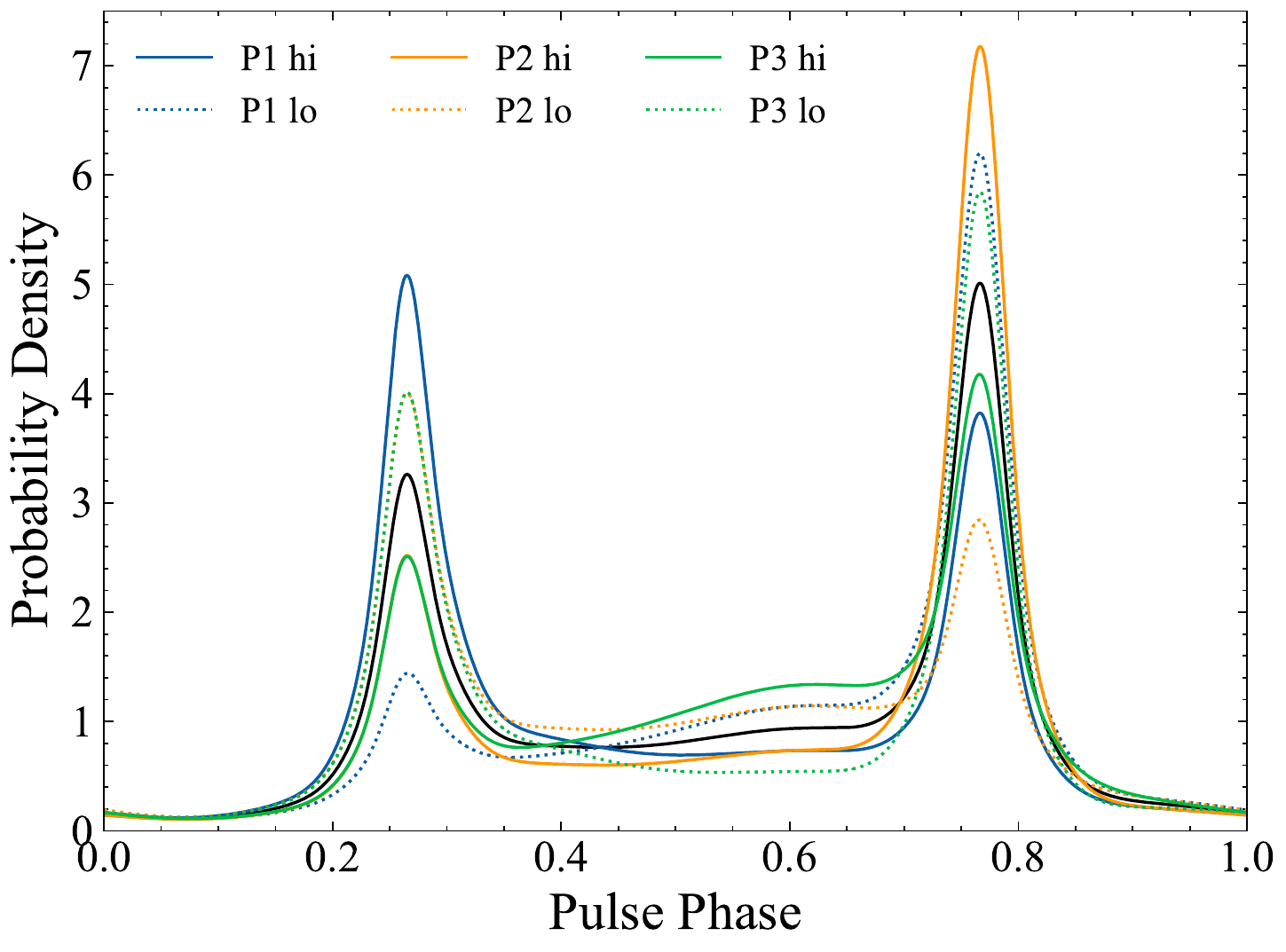}
\caption{\label{fig:geminga_basis}As Figure \ref{fig:vela_basis} for Geminga, but note different phase range.}
\end{figure}

We now seek to quantify the largest shape variations the data can
accommodate.  In doing so, we are guided by the results from
\S\ref{sec:amplitude} and the empirical analysis:  the time-averaged
distribution of photon pairs is very similar to $f(\phi)$, and if
pulse shape variations are accompanied by intensity variations, they
cannot be much larger than 20\%.  We also seek a physically motivated
model: although the $\gamma$-ray emission contributing to a single
spin phase can be accumulated from a large portion of the
magnetosphere through caustic-like effects, the portions of the
magnetosphere contributing to, e.g., P1, P2, and the bridge/P3 in Vela
are distinct.  Thus, considering changes localized to pulse profile
features will target large-scale but distinct regions of the
magnetosphere.

We thus adopt a family of $N_b$ basis functions, $b_i({\phi}$), which
describe the variation in the pulse profile from rotation to rotation.
Over many rotations, the individual pulse profiles must yield the
long-term light curve, so by construction $\sum_i
b_i({\phi})/N_b=f(\phi)$.  For both Vela and Geminga, with
well-defined two-peaked light curves and bridge emission, we choose
$N_b=6$ corresponding to an increment and decrement of each of the
three main features.  The basis functions are defined via a strength
parameter $\alpha$, with each basis function reducing to the null
hypothesis: $b_i(\alpha,\phi)\rightarrow f(\phi)$ as
$\alpha\rightarrow0$.  Figure \ref{fig:vela_basis}
(\ref{fig:geminga_basis}) illustrates the basis functions for Vela
(Geminga) with $\alpha=0.15$ (15\%) variations.  Even this modest
total variation results in changes of the relative peak amplitudes by
about 2.  And indeed, $\alpha$ cannot exceed $\sim$0.2 as
this results in the removal of gross features of the light curve and a
negative pdf.  This is not a significant restriction since
$\alpha=0.2$ already captures very large shape variations, and it
ensures that the models tested in this analysis are consistent with
the amplitude variation limits established above.

\begin{figure}
\centering
\includegraphics[angle=0,width=0.98\linewidth]{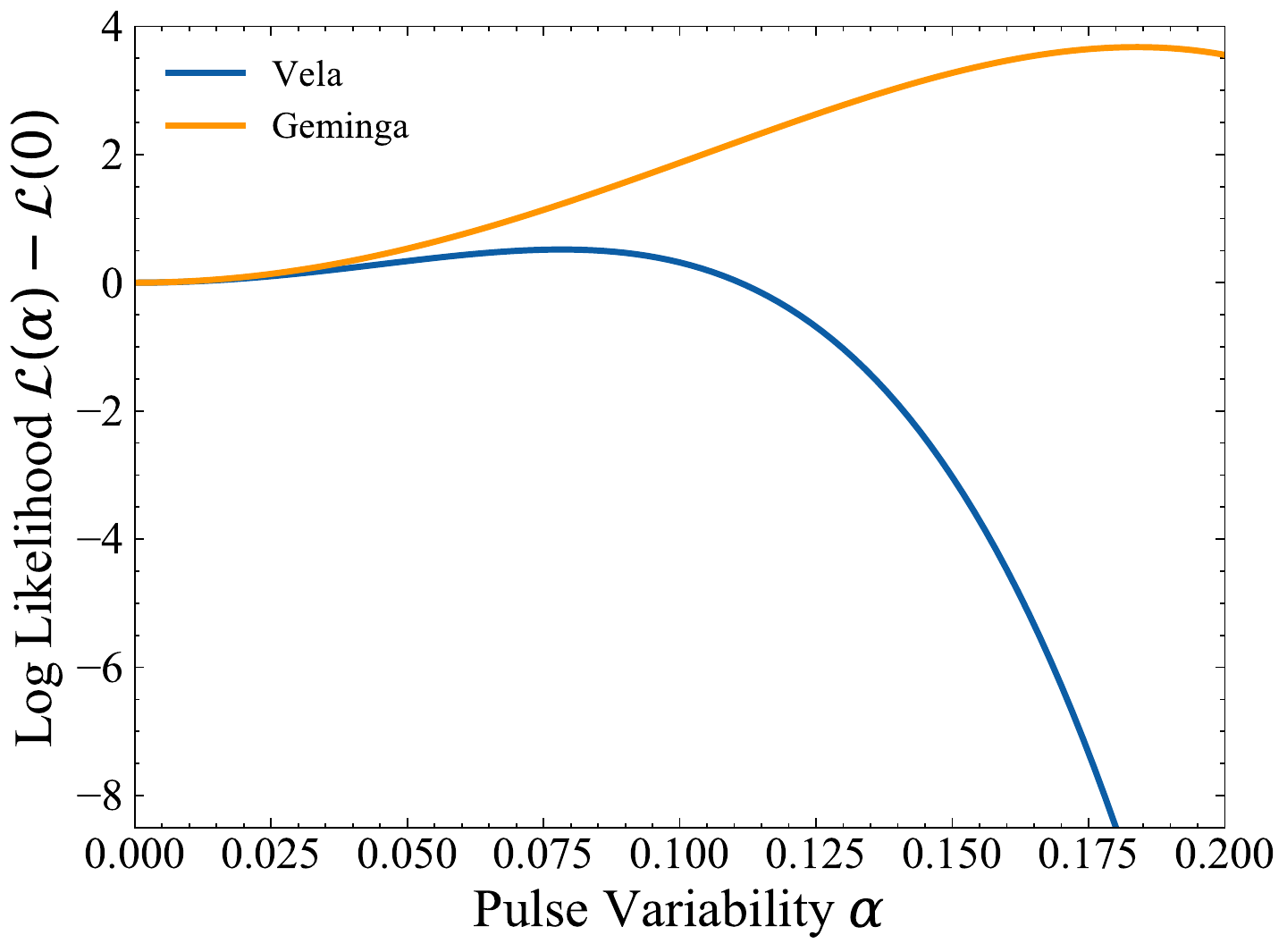}
\caption{\label{fig:vela_alpha_logl}The log likelihood difference for
Vela and Geminga for pulse shape variability characterized by a basis with
strength $\alpha$ (see main text) relative to the null hypothesis.}
\end{figure}

We proceed via maximum likelihood.  On a given rotation, the photons
are emitted according to one of the basis functions.  Considering just
a single pair of photons, the probability density function is then
\begin{align}
&(1-w_{1})(1-w_{2}) + \sum_i \pi_i
\big[ w_{1}(1-w_{2})b_i(\alpha,\phi_{1}) + \\
\nonumber & w_{2}(1-w_{1})b_i(\alpha,\phi_{2}) +
w_{1}w_{2}
b_{i}(\alpha,\phi_{1})b_{i}(\alpha,\phi_{2})\big],
\end{align}
where the $\pi_i$ describe the probability of the $i$th basis function
as being the correct one for the rotation in question.  For
convergence to the time-averaged profile, $<\pi_i>=1/N_b$.  We have no \textit{a priori} knowledge of $\pi_i$ for any given pulse, so we let it assume the time-averaged value $\pi=1/6$, allowing direct evaluation of the likelihood as a function of $\alpha$.  The
resulting log likelihood surface is shown in Figure
\ref{fig:vela_alpha_logl}.  The likelihood peak for Vela \changed{is
near $\alpha=0.08$},
but the significance is consistent with $\alpha=0$.  Consequently, we
assume a uniform prior on $\alpha$ and estimate a 95\% confidence
upper limit \changed{of $\alpha<0.12$.}

Next, we consider the empirical distribution function for Geminga
(Figure \ref{fig:geminga_empirical_dphi}).  As with Vela, it appears
to largely follow the null hypothesis.  The projected distribution in
$\phi_2-\phi_1$ does show a small excess near the origin, indicating a
possible small correlation for photons to arrive nearby in phase.
Using the same basis-based technique (see Figure
\ref{fig:geminga_basis} for the basis functions for Geminga), we find
the maximum likelihood value occuring around $\alpha=0.18$,
\changed{with $\delta\log\mathcal{L}=3.67$}, near the edge of the validity of our
model.  Because this model satisfies the conditions of Wilks' Theorem,
we can quickly estimate the significance of the alternative hypothesis
as \changed{99.3\%}.  Thus there is modest evidence for pulse shape variability
for Geminga, with an amplitude consistent with the largest allowed
intensity variations.  By skipping marginalization, we can determine
the likelihood-maximizing basis function for each pair and thus obtain
the distribution of preferred bases.  We find bases which modify P1
and P2 are most prevalent, with P3 ``hi'' being least preferred.
Thus, if the pulse shape variation is real, it is concentrated in the
peaks, consistent with the empirical analysis above.  However, we
caution that this result strongly depends on our assumption of a
particular basis.

\section{Multi-rotation correlations}
\label{sec:multirot}

\begin{figure}
\centering
\includegraphics[angle=0,width=0.98\linewidth]{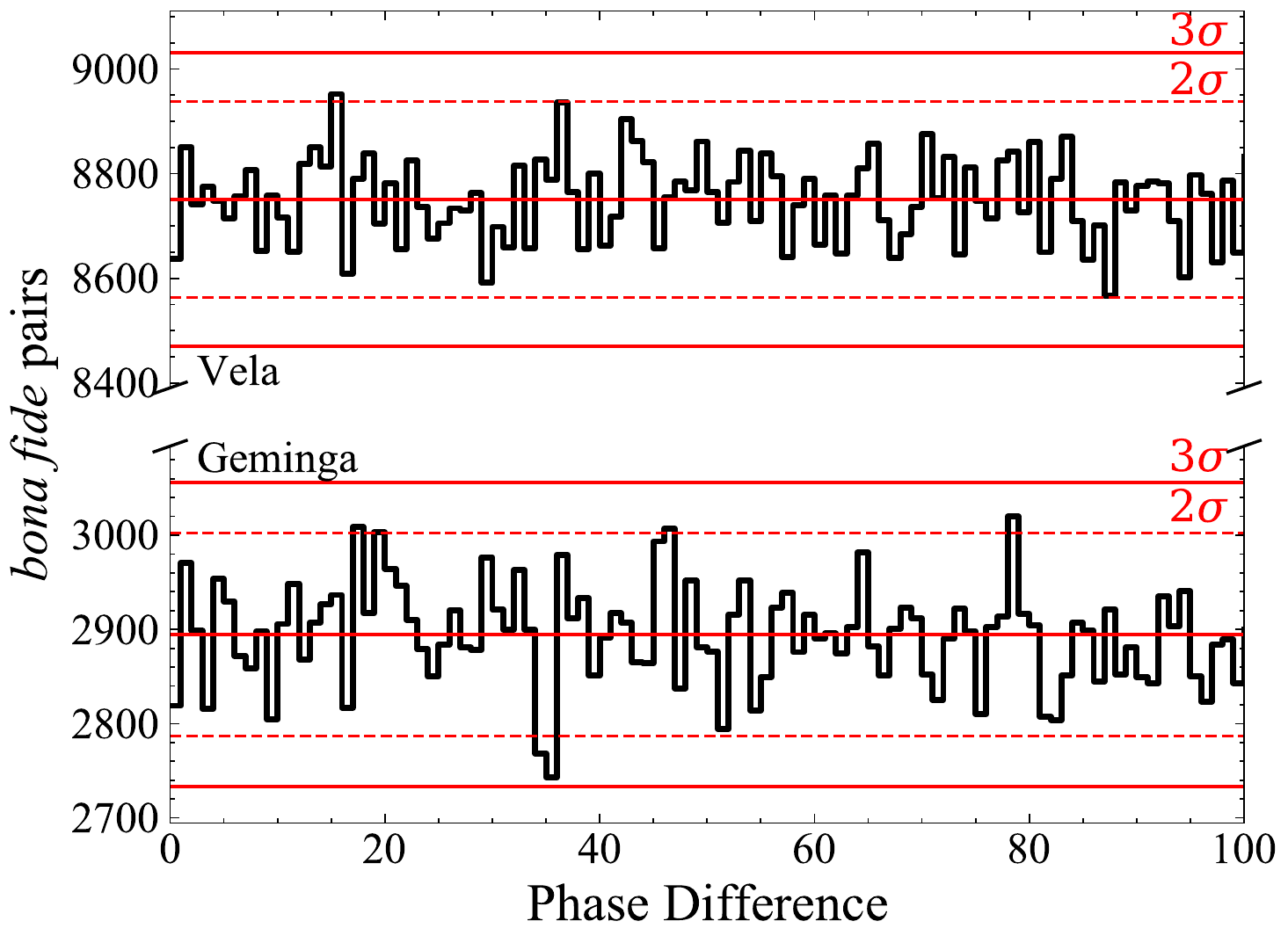}
\caption{\label{fig:joint_multirot_amp}The observed \textit{bona fide}
photon pairs for Vela and Geminga, binned to integer phase differences.  The
expected number of pairs, as calculated by Monte Carlo siulations, and
the 2- and 3-$\sigma$ uncertainties, are
shown by horizontal lines.  The data are in good agreement with an
absence of amplitude fluctuations both \textit{in toto} and for any
particular phase separation.}
\end{figure}

We have also considered correlations on time scales greater than one
neutron star rotation, which could be expected e.g. from
magnetospheric state switching on longer time scales.  We carried out
a similar procedure, identifying photons with the requisite phase
separation and using their weights to calculate the probability that
they both originate from the pulsar.  Here, we apply a simpler
criterion, selecting as ``pairs'' those photons whose phase separation
lies below a given threshold, say $\delta\phi<\Phi$.  A threshold of
$\Phi=0.5$ approximately reproduces the single-rotation selection
above, while $\Phi=1$ produces about twice as many pairs, because it
has contributions from two pulses.

\begin{figure}
\centering
\includegraphics[angle=0,width=0.98\linewidth]{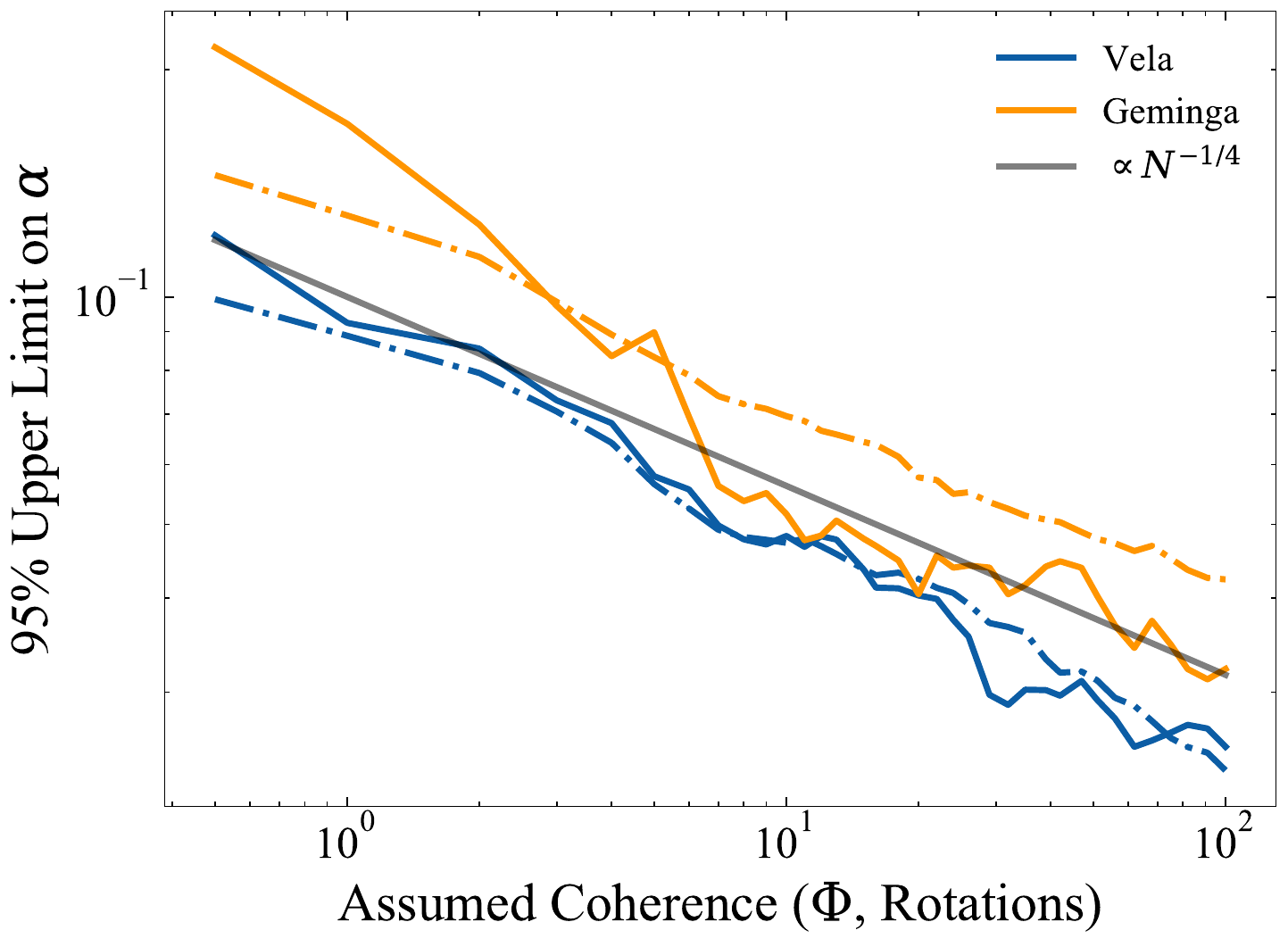}
\caption{\label{fig:multirot_shape}Limits on the strength of the
variability parameter $\alpha$ (see main text) when including pairs
collected over the indicated number of rotations.  The solid lines
give the results for the data, while dashed lines give the average of
10 Monte Carlo realizations in which any correlations in the real data
have been broken.  The scaling in both cases is in good agreement with
the total number of pairs to the one-fourth power.}
\end{figure}

As before, we examine amplitude fluctuations by studying the total
number of pairs up to an assumed coherence of $\Phi=100$ rotations,
with results shown in Figure \ref{fig:joint_multirot_amp} for the two
bright pulsars.  Counting pairs based on exposure is more difficult in
this case, because each rotation of the pulsar is no longer
statistically independent.  Thus, to calibrate the mean expectation,
we use a simple Monte Carlo approach in which we permute the integer
phase of each photon.  This preserves the overall exposure pattern but
breaks any short-term correlations.  The resulting predictions, along
with the corresponding Poisson uncertainties, are drawn in the figure.
Because each of the observed pair counts lies within this band, we
conclude there is no evidence for amplitude variations.  By assuming
longer coherence scale, the number of pairs increases substantially,
and the limits on amplitude fluctuations tighten.  The scaling
simply depends on the total number of pairs, thus on $\Phi$, and we
arrive at the limits on relative amplitude fluctuation for an assumed
coherence time $\Phi$ in neutron star rotations:
\changed{$<21\%(\Phi/0.5)^{-1/2}$ for Vela and $<16\%(\Phi/0.5)^{-1/2}$ for
Geminga.  For $\Phi=10$ this is 4.7\% and 3.6\%, and for $\Phi=100$,
1.5\% and 1.1\%.}

Likewise, we can feed this larger population of pairs into the pulse
shape analysis of \S\ref{sec:pulse_shape}.  Here, because of the
marginalization over the unknown basis for each pulse pair, the
scaling with the number of pulses is less straightforward, and we have
carried out the analysis directly on the data for the increasingly
larger populations of pairs.  The results, as a function of assumed
coherence length $\Phi$, are shown in Figure \ref{fig:multirot_shape}.
N.B. that the modest evidence for pulse shape variability in
Geminga detected in \S\ref{sec:pulse_shape} drops as soon as
additional rotations are added and the limit quickly asymptotes to the
same dependence as for Vela.  For $\Phi=10$, the 95\% limit on the
shape variation parameter $\alpha$ is about 5\% for both pulsars, and
for $\Phi=100$ about 3\%.  The limits on the parameter seem to scale
as the total observed pairs to the one-fourth power, rather than the
one-half power as expected for Poisson uncertainties.


\section{Other Bright Pulsars}
\label{sec:other}

\begin{table}
\centering
\begin{tabular}{l | r | r | r | r}
Type & Obs. & Chance  & \textit{bona fide}  & Pred.\\
\hline
\hline
  J0534+2200 events & 695,267 & --- & 398,271.7 & --- \\
  J0534+2200 pairs & 125 & 85.5 & 39.5 & 35.6 \\
  J1709-4429 events & 1,177,343 & --- & 359,382.0 & --- \\
  J1709-4429 pairs & 918 & 828.1 & 89.9 & 84.8 \\
  J1057-5226 events & 283,529 & --- & 90,115.8 & --- \\
  J1057-5226 pairs & 113 & 103.8 & 9.2 & 10.0 \\

\hline
\end{tabular} 
\caption{\label{tab:counts_other}As Table \ref{tab:counts}, for
fainter pulsars.}
\end{table}

The Crab pulsar (J0534$+$2200) is an interesting target due
to its known giant pulse phenomenon and the recent discovery of
enhanced X-ray emission associated with such giant pulses
\citep{Enoto21}.  However, its fast spin frequency yields a typical
weighted photon rate per rotation of only $1.5\times10^{-4}$, and
consequently we have observed only 39.5 \textit{bona fide} pairs, in
reasonably good agreement with the null hypothesis prediction of 35.6.
These are simply too few pairs to probe single pulse variations beyond
this simple sanity check.

Another ``EGRET'' pulsar PSR~J1709$-$4429 is bright but embedded deep
in the Galactic plane, giving a relatively high background rate,  and
consequently only 10\% of the observed pairs are likely to be pulsed.
It has a median weighted photon rate of $4.7\times10^{-4}$ and
produces 89.9 \textit{bona fide} pairs compared to a prediction of
84.8, in excellent agreement.  Finally, PSR~J1057$-$5226 has a mean
weighted photon rate of $2.2\times10^{-4}$, and its exposure pattern
leads to mean predicted and observed pair rates of 10.0 and 9.2,
respectively.  The next few brightest LAT pulsars all also have faster
spin periods and consequently even fewer pairs.

\section{Discussion}
\label{sec:discussion}




Thus, for at least two pulsars, we can restrict variations in
amplitude and pulse shape to $\leq$20\%. (Though we caution that with
our definition, shape variations of 20\% still encapsulate 
substantial changes to the pulse profile.)  These constraints allow us
to rule out large, coherent variations of the full $\gamma$-ray
emitting volume.

Consider the case of a nearly aligned rotator, with
$\gamma$-rays emitted within $\sim$$R_{LC}$ (light cylinder radius) of
the separatrix.  Kink instabilities within the ECS observed in PIC
simulations can deform this region substantially, producing deviations
of 0.1\,rad or more, although the amplitudes are likely to be smaller
in magnetospheres with realistic magnetic fields \citep{Cerutti16}.  Beaming of the $\gamma$-ray emission would then
result in either removal of the flux from the line-of-sight or a shift
to a different spin phase, either of which could be detected with this
method.  Thus, either the amplitude of such instabilities must be
smaller, or else the $\gamma$-ray emission must arise on different
spatial scales.

As discussed in \S\ref{sec:intro}, one possibility is
that magnetic reconnection in the ECS could fragment into many small
islands which damp large-scale instabilities.  Another possibility is
that the $\gamma$-ray emission instead arises over many $R_{LC}$, i.e.
within the striped wind \citep{Petri12}, such that material
corresponding to many neutron star rotations contributes to an
apparent pulse, and the observational effects of instability average
away.  Intriguingly, \citet{An20} detected orbital modulation of
pulsed emission from a millisecond pulsar, and one possible way to
arrange this is via inverse-Compton scattering of light from the
companion by $\gamma$-ray emitting particles in the striped wind.  However, this process may be inefficient and instead such modulation may rely critically on the presence of the companion star \citep{Clark21}.

We can also rule out wholesale state changes between FF and
electrosphere configurations.  Such state changes seem to increase as
a pulsar ages, and are potentially related to the pair multiplicity:
\citet{Chen20} note that the duty cycle of these oscillations is
related to the time it takes for the global current to deplete the
charge cloud formed when the pulsar is in the active (ECS present)
state.  Thus, oscillation could operate on a range of timescales, from
one period to many periods.  Using the results of
\S\ref{sec:multirot}, we can limit the time spent in such ``dead''
states to less than a few percent.  Moreover, the variability models
for both shape and amplitude did not include the extreme case of
nulling, so the constraints are even stronger.  Comparison to behavior
found by, e.g. \citet{Chen14} in which the magnetosphere ``breathes''
at the spin period, varying the position of the Y-point, must await
more quantitative predictions from simulation.

The lack of observable state changes is perhaps not surprising for
Vela and Geminga, energetic pulsars whose bright $\gamma$-ray
luminosity requires the formation of many $e^+$/$e^-$ pairs.  However,
such changes have been observed in other pulsars over a wide range of
time scales.  In the most relevant case, substantial $\gamma$-ray
pulse profile and flux changes, with correlated spin down changes,
have been observed in the energetic, radio-quiet, Geminga analogue
PSR~J2021$+$4026 \citep{Allafort13} with an apparent variability
timescale of about 3 years.  \citet{Kerr16} found evidence for
quasi-periodic modulation of the spindown rate with 1--3 year
timescale in a sample of energetic, radio-loud pulsars.  In the same
sample of high spindown-power radio pulsars---thus likely $\gamma$-ray
emitters---\citet{Brook16} found additional evidence of aperiodic,
year-scale correlations between radio profile variability and spindown
rate.  Such timescales are difficult to account for with any
magnetospheric source of pulsar timing noise, and are more reminiscent
of older, less energetic radio pulsars, which demonstrate long term
nulling with decreased spindown rate \citep[e.g.][]{Kramer06} but also
less dramatic correlated spindown-rate and pulse profile changes on
shorter timescales \citep{Lyne10}.

An unusual magnetic configuration may be required for such strong
observable changes.  J2021$+$4026 may be nearly aligned and viewed
near the spin equator, accounting both for its lack of radio emission
and very broad (much of its emission is unpulsed) $\gamma$-ray
profile.  Further, \citet{Philippov15a} note that more oblique pulsars
have a suppressed kink instability.  And interestingly, on even
shorter (hour) timescales, \citet{Hermsen13} directly observed changes
in the open and closed field zones via correlated radio and X-ray
emission for PSR~B0943$+$10, which is also thought to be nearly
aligned.  However, the discovery of a nearly-orthogonal rotator with
similar radio/X-ray correlated mode changes \citep{Hermsen18}
precludes any simple criteria relating metastable equilibria and
magnetic field alignment.

From the standpoint of simulations, much additional work is needed to
make such observational features accessible.  A typical PIC simulation
lasts only about a few neutron star rotations, so measuring
instability properties robustly---e.g. fluctuation spectra of the ECS
deformation---will require a substantially larger investment of
computational resources.  \changed{\citet{CeyhunAndac22} recently took
a step in this direction, using two-dimensional PIC simulations to analyze
variability in pulse flux over tens of rotations.  The resulting
distribution of ``subpulse'' fluxes may be in tension with the results
reported here.}

Extending this variability analysis to more $\gamma$-ray pulsars could
reveal more rare cases like J2021$+$4026 and thus better connect with
simulations and multiwavelength observations, but this is
unlikely.  Analyses on such short timescales were not envisaged at
all prior to the launch of \textit{Fermi}, and are only possible
thanks to improvements in the event reconstruction \citep{Atwood13}
and the long dataset. Together these yield enough pulsed pairs to
provide meaningful constraints for the two brightest pulsars.  And
indeed additional LAT data (another $\sim$10 years) will improve these
constraints by a few tens of per cent, which may be enough to
determine if the modest evidence for Geminga pulse shape variations
grows into a real detection.  However, this improvement is not good
enough to increase the sample of useful pulsars beyond two.

A 10$\times$ sensitivity improvement---the canonical threshold for a
new experiment---would make measurements for dozens more young pulsars
accessible.  (Single-pulse studies of millisecond pulsars would
require about 1000$\times$ improved sensitivity!)  But such an
observatory would require a much larger payload than the LAT, which is
already fairly efficient relative to its geometric cross section,
and with more complete $\gamma$ conversion requiring either thicker
tungsten foils (degrading the PSF) or a much deeper tracker
(decreasing the field-of-view).  This larger instrument could be made
about 30\% lighter per unit area by sacrificing $>$10\,GeV performance
with a thinner calorimeter that still fully contains 1\,GeV showers.
For a fixed geometric area, an improved PSF would help substantially
for pulsars embedded in the Galactic plane.  E.g., while for Vela and
Geminga, the number of pairs is ``signal dominated''---more come from
the pulsar than from the background---this is not at all true for
PSR~J1709$-$4429, where only about 10\% of the pairs are \textit{bona
fide}.  In this case, improving the angular resolution by 2$\times$
(reducing the background by $\sim$4$\times$) is about as effective as
boosting the effective area by a factor of 3.  Thus, a ``super LAT''
four times larger than LAT and with an improved PSF could meet this
10$\times$ threshold.  Such a configuration would also maintain the
squat configuration preferred for a large field of view because the
additional depth needed to achieve the better PSF would be offset by
the larger area.  Unfortunately no such instrument is under
development.



\acknowledgements
The \textit{Fermi} LAT Collaboration acknowledges generous ongoing
support from a number of agencies and institutes that have supported
both the development and the operation of the LAT as well as
scientific data analysis.  These include the National Aeronautics and
Space Administration and the Department of Energy in the United
States, the Commissariat \`a l'Energie Atomique and the Centre
National de la Recherche Scientifique / Institut National de Physique
Nucl\'eaire et de Physique des Particules in France, the Agenzia
Spaziale Italiana and the Istituto Nazionale di Fisica Nucleare in
Italy, the Ministry of Education, Culture, Sports, Science and
Technology (MEXT), High Energy Accelerator Research Organization (KEK)
and Japan Aerospace Exploration Agency (JAXA) in Japan, and the
K.~A.~Wallenberg Foundation, the Swedish Research Council and the
Swedish National Space Board in Sweden.
 
Additional support for science analysis during the operations phase
is gratefully acknowledged from the Istituto Nazionale di Astrofisica
in Italy and the Centre National d'\'Etudes Spatiales in France. This
work performed in part under DOE Contract DE-AC02-76SF00515.

Work at NRL is supported by NASA.  The author is grateful to David
Smith, Jean Ballet, and Zorawar Wadiasingh for helpful suggestions, to
Alice Harding for a careful review of the manuscript, \changed{and to
the anonymous journal referee, whose suggestions improved this work.}

\facilities{Fermi}

\bibliographystyle{aasjournal}
\bibliography{sr}


\end{document}